# Radiation from a charged particle-in-flight from a laminated medium to vacuum

**L Sh Grigoryan**[1,3], **A R Mkrtchyan**[1], **H F Khachatryan**[1], **S R Arzumanyan**[1] and **W Wagner**[2]

[1]Institute of Applied Problems in Physics, 25 Hr. Nersessian Str.,0014 Yerevan, Armenia
[2]Forschungszentrum Dresden-Rossendorf, Institute of Radiation Physics, POB 510119, 01328 Dresden, Germany

E-mail: levonshg@mail.ru

**Abstract.** The radiation from a charged particle-in-flight from a semi-infinite laminated medium to vacuum and back,- from vacuum to the laminated medium, has been investigated. Expressions for the spectral-angular distribution of radiation energy in vacuum (at large distances from the boundary of laminated medium) were obtained for both the cases with no limitations on the amplitude and variation profile of the laminated medium permittivity. The results of appropriate numerical calculations are presented and possible applications of the obtained results are discussed.

## 1. Introduction

The period of existing periodical structures varies from ~ $10^{-8}$ (crystals) to microns (photonic crystals) and up to ~ 1 *cm* (undulators). It is natural that the radiation from charged particles in the above periodical structures has multiple manifestations that are of interest for practical applications [1, 2].

For instance, it is specific that in the periodic structures there are forbidden bands on dispersion curves for electromagnetic waves (see, e.g. [3]). The presence of such bands shall influence the radiation from a charged particle in periodical structures and for this reason these may serve as additional means for controlling the particle radiation parameters. This mechanism was first investigated in the case of an infinite laminated medium composed of two alternating plates having different values of permittivity $\varepsilon$ [4]. In [5, 6] the infinite stack of plates was replaced with a laminated medium, the permittivity of which is an arbitrary one-dimensional periodic function (the problem of a stack of the finite number of plates has been investigated in [7], see also [8]). However, the calculation methods used in [5,6] (quasi-classical approximation, perturbation theory) failed to allow for the influence of forbidden bands on the emitted electromagnetic waves. The present paper deals with the solution of this problem.

The problem is formulated in section 2. Main stages of analytical calculations are given in section 3, the results of numerical calculations are analyzed in section 4 and the main results of the paper are summarized in the last section.

---

[3] To whom any correspondence should be addressed.

## 2. Formulation of the problem

Consider the motion of charge $q$ along the positive direction of OZ axis (see figure 1a) at its travel from a semi-infinite laminated medium ($z < 0$, range 1) to vacuum ($z > 0$, range 2). The plane $z = 0$ is taken as an interface between the laminated medium with vacuum.

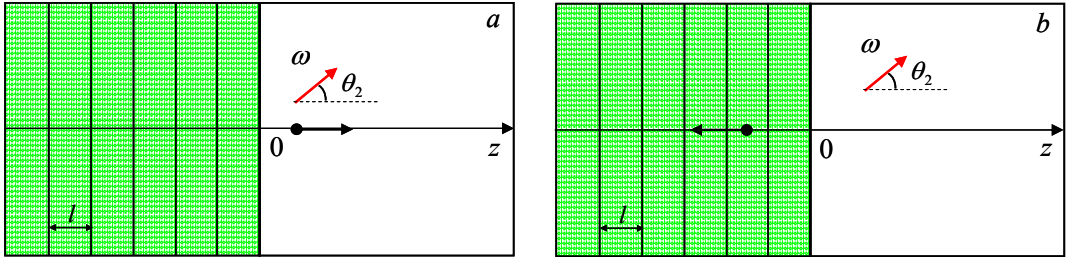

**Figure 1.** The particle emerging from the semi-infinite laminated medium to vacuum (*a*) and vice versa, – incoming from vacuum to the laminated medium (*b*).

We shall assume that the electrical permittivity $\varepsilon_1$ and magnetic permeability $\mu_1$ of the laminated medium are independent of $(x,y) \equiv \vec{\chi}$ and periodically vary along the direction of particle motion:

$$\varepsilon_1(z-l) = \varepsilon_1(z), \qquad \mu_1(z-l) = \mu_1(z), \qquad (1)$$

with arbitrary profiles ($l$ is the period of laminated medium). As a result, the electrical permittivity $\varepsilon$ and magnetic permeability $\mu$ of the system on the whole may be written as

$$\varepsilon(z) = \varepsilon_1(z), \qquad \mu(z) = \mu_1(z) \qquad \text{for } z < 0 \quad \text{(range 1)}$$
$$\varepsilon(z) = 1, \qquad \mu(z) = 1 \qquad \text{for } z > 0 \quad \text{(range 2)}. \qquad (2)$$

If we represent the energy $W_2$ of radiation transmitted through the surface $z = z_0$ distant from $z = 0$ in the range 2 as

$$W_2 = \int I_2(\omega, \theta_2) d\omega d\theta_2 \qquad (3)$$

during the whole time of particle motion, then $I_2(\omega, \theta_2)$ shall describe the spectral-angular distribution of the energy of radiation that propagates in vacuum along the direction of particle motion (forward radiation). We aim at an analysis of this quantity.

Note that at the reversal of the direction of particle motion v $\to$ -v (see figure 1b), we arrive at the problem of radiation from particle incoming into the laminated medium. In this case $I_{2,v \to -v}(\omega, \theta_2)$ shall describe the radiation in vacuum, propagating in the backward direction to the particle motion (backward radiation).

## 3. The stages of analytical calculations and the final formula

Taking into account the azimuthal symmetry of problem and using a Fourier transform

$$f_{u\omega}(z) = \frac{1}{(2\pi)^3} \int f(\vec{r},t) \exp[i(\omega t - \vec{u}\vec{\chi})] d\vec{\chi} dt \qquad (4)$$

the solution of the set of Maxwell equations is reduced to the solution of a single equation [4]

$$[\varepsilon \frac{\partial}{\partial z}\left(\frac{1}{\varepsilon}\frac{\partial}{\partial z}\right) + \frac{\omega^2}{c^2}\varepsilon\mu - u^2]\varepsilon E_{u\omega} = 4\pi[\varepsilon \frac{\partial}{\partial z}\left(\frac{\rho_{u\omega}}{\varepsilon}\right) - i\frac{\omega}{c^2}\varepsilon\mu j_{u\omega}] \qquad (5)$$

(the basic equation of the problem under consideration). In this equation $\varepsilon \equiv \varepsilon_\omega$ and $\mu \equiv \mu_\omega$ are Fourier transformations of the electric and magnetic permittivities (taking into account the frequency dispersion, the spatial dispersion being assumed negligibly small). Here we assume that $\varepsilon_1 = \varepsilon'_{1\omega} + i\varepsilon''_{1\omega}$ and $\mu_1 = \mu'_{1\omega} + i\mu''_{1\omega}$ are complex valued functions (making allowance for the absorption of radiation by the material of laminated medium). Having determined the solution $E_{u\omega}$ of the equation (5), one may calculate the strengths of electric and magnetic fields by means of the following formula:

$$\vec{E}_{u\omega} = E_{u\omega}\vec{n}_z + \frac{i\vec{u}}{\varepsilon u^2}\left(\frac{\partial}{\partial z}\varepsilon E_{u\omega} - 4\pi\rho_{u\omega}\right), \qquad \vec{H}_{u\omega} = \frac{\omega}{cu^2}(\varepsilon E_{u\omega} + 4\pi\frac{i}{\omega}j_{u\omega})\vec{u}\times\vec{n}_z. \qquad (6)$$

Here

$$\rho(\vec{r},t) = q\delta(x)\delta(y)\delta(z-\mathrm{v}t) \quad \text{and} \quad \mathrm{v}\rho_{u\omega}(z) = \frac{q}{(2\pi)^3}\exp(i\frac{\omega}{\mathrm{v}}z) = j_{u\omega}(z) \qquad (7)$$

is the density of current connected with the charged particle.

In range 2 equation (5) shall have the following solution:

$$E_{u\omega} = E_{u\omega}^q + \frac{iq}{2\pi^2\omega}a_2\exp(ik_2z), \qquad k_2 = \sqrt{\omega^2/c^2 - u^2}, \qquad (8)$$

where the first term is the known field of charge in vacuum [6], and the 2nd one describes the free field (the radiation field). At large distances from the semi-infinite medium the spectral-angular distribution of radiation energy is determined by the well-known expression [6,7]

$$I_2(\omega,\theta_2) = \frac{2q^2}{\pi c}\frac{\cos^2\theta_2}{\sin\theta_2}|a_2(\omega,u)|^2, \qquad \text{where} \quad u = \frac{\omega}{c}\sin\theta_2, \qquad (9)$$

and $\theta_2$ is the angle showing the direction of radiation with respect to OZ axis (see figures 1a,b). For calculation of $a_2$ it would be enough to equate (8) to the complete solution of equation (5) that is valid for all $-\infty < z < \infty$. This solution is determined most easier by means of the method of Green functions, because the Green function is the solution of a homogeneous equation, which is obtained from (5) by means of substitution $\rho_{u\omega} = j_{u\omega} = 0$. In this case the general solution of homogeneous equation inside the laminated medium may be represented according to the Bloch theorem (see, e.g. [9]) as a superposition $b_1 L_+(z) + c_1 L_-(z)$ of running waves $\exp(\pm ik_1 z)$ modulated with the period of laminated medium, i.e.,

$$L_\pm(z) = w_\pm(z)\exp(\pm ik_1 z), \quad \text{where} \quad w_\pm(z-l) = w_\pm(z). \qquad (10)$$

Here

$$\cos k_1 l = \frac{\chi_a(\tau)\dot\chi_b(0) + \chi_a(0)\dot\chi_b(\tau) - [\dot\chi_a(\tau)\chi_b(0) + \dot\chi_a(0)\chi_b(\tau)]}{2[\chi_a(0)\dot\chi_b(0) - \dot\chi_a(0)\chi_b(0)]} \qquad (11)$$

and

$$L_\pm(z) = [\chi_b(0) - \exp(\pm ik_1 l)\chi_b(\tau)]\chi_a(z) - [\chi_a(0) - \exp(\pm ik_1 l)\chi_a(\tau)]\chi_b(z), \qquad (12)$$

where $\chi_a(z)$ and $\chi_b(z)$ is the pair of any linearly independent solutions of the homogeneous equation, $\tau = -l$, and the point over the function means the differentiation with respect to its argument. The quasi-wave number $k_1(\omega,u)$ is determined from (11) with an accuracy of the sign. To fix the sign we shall assume that

$$k_1 = k_1' + ik_1'', \qquad \text{where} \quad \omega k_1''(\omega,u) \geq 0 \qquad (13)$$

($k_1'' \neq 0$ as the absorption of radiation in the material of laminated medium was taken into account). The forbidden bands of the laminated medium mentioned in the Introduction correspond to the values of $\omega,u$, for which $\omega k_1''(\omega,u)$ stays positive even when $\varepsilon_1''/\varepsilon_1' \to 0$. The electromagnetic oscillations with these values of $\omega,u$ can not freely propagate in a laminated medium as the corresponding solutions $L_\pm(z)$ of the homogeneous equation exponentially decrease along one of the directions of OZ axis.

Omitting the intermediate calculations we give the final expression

$$a_2 = \frac{\omega}{2k_2 v}\left\{\frac{1-v^2/c^2}{1-k_2 v/\omega} - \frac{1-v^2/c^2}{1+k_2 v/\omega}\cdot\frac{k-\varepsilon_1 k_2}{k+\varepsilon_1 k_2} \right.$$
$$\left. -\left[\varepsilon_1 - 1 + \frac{\sigma}{1-\exp[i(k_1-\omega/v)l]}\right]\frac{2k_2}{k+\varepsilon_1 k_2}\right\}_{z=0}. \quad (14)$$

Entering into this expression are the values of $\varepsilon_1(z)$ and $k(z)$ in point $z = 0$,

$$k = i\frac{\dot{L}_-(0)}{L_-(0)} \quad \text{and} \quad \sigma = i\frac{\omega\varepsilon_1(0)}{vL_-(0)}\int_{-l}^{0}\left(1-\varepsilon_1\mu_1\frac{v^2}{c^2} + \frac{iv\dot{\varepsilon}_1}{\omega\varepsilon_1}\right)\frac{L_-}{\varepsilon_1}\exp(i\frac{\omega}{v}z)\,dz. \quad (15)$$

Three terms in (14) correspond to three types of waves: the 1st term corresponds to the wave emitted in vacuum in the forward direction, the 2nd term – to the wave emitted in vacuum primarily in backward direction and then reflected from the interface between vacuum and the laminated medium, and the 3rd term – to the perturbation emitted in the laminated medium in the forward direction and traversed through the interface between the laminated medium and vacuum. In the partial case of $\varepsilon_1(z)$, $\mu_1(z) = const$ in (14), there follows the well known expression for transition radiation at the passage of particle from a homogeneous medium to vacuum [5-7].

To obtain the expressions for radiation in the backward direction (see figure 1b) one has to replace v by - v in above formulae.

## 4. Numerical results
We confine ourselves to the consideration of a model of laminated medium (see figure 2) with non-harmonic variations of $\varepsilon_1(z)$ and large modulation depth $\Delta\varepsilon_1' \sim \overline{\varepsilon_1'(z)}$ so, that

$$\Delta\varepsilon_1' = 1, \qquad \varepsilon_1'(0) = \overline{\varepsilon_1'(z)} = 1.5, \qquad \varepsilon_1''/\varepsilon_1' = 0.01 \quad \text{and} \quad \mu_1 = 1. \quad (16)$$

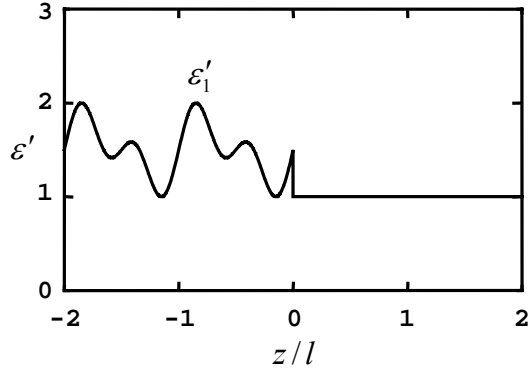

**Figure 2.** The model of $z$-dependence of the real part of electric permittivity $\varepsilon(z)$ of the system. The laminated medium with $\varepsilon = \varepsilon_1$ is to the left and vacuum is to the right of $z = 0$.

It was assumed that in the range of $\omega$ under consideration one can neglect the dispersion of electromagnetic waves in the laminated medium. If we take the energy of particle (electron) equal to $30 MeV$, then the Cherenkov condition in the form

$$\cos\overline{\theta}_1 = c/v\sqrt{\overline{\varepsilon_1'}} \quad (17)$$

is satisfied. Here the value $\overline{\theta}_1 \cong 35°$ specifies the characteristic direction of Cherenkov radiation (CR) in the laminated medium. Now introduce a dimensionless parameter

$$\omega_l = \omega l / \pi v. \qquad (18)$$

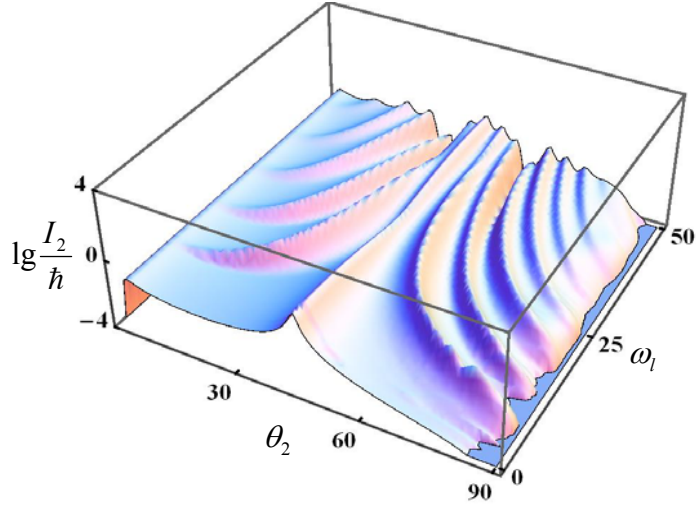

**Figure 3.** Spectral-angular distribution of radiation from $30 MeV$ electron emerging from the laminated medium with non-harmonic variations of $\varepsilon_1(z)$ (see figure 2, (16)) to vacuum (radiation in the forward direction).

The spectral-angular distribution $I_2(\omega, \theta_2)$ of the energy of particle radiation is shown in figure 3. The central part of this figure with
$$\theta_2 \approx \arcsin(\sqrt{\varepsilon_1'} \sin \overline{\theta}_1) \cong 45° \qquad (19)$$
corresponds to the main beam of particle CR. On either sides of this central part the lines of diffracted radiation split up (the beams of the 1st, 2nd and higher orders of parametric radiation). The results of our numerical calculations indicate that the distribution $I_2(\omega, \theta_2)$ on a separate branch is determined by the amplitude and profile of $\varepsilon_1$ variations.

It is noteworthy that some part of the branches of spectral-angular distribution reaches the range of $\theta_2 < 10°$ and superimposes on the existing small peak in the angular distribution of radiation in this range. It is clearly seen in the lower left angle of figure 3 and corresponds to the transition radiation (TR) of particle at the interface of laminated medium with vacuum. Thus, the peaks from diffracted beams of parametric radiation superimpose on the spectral distribution of TR.

Now consider the case of particle flying into the laminated medium from vacuum (backward radiation in figure 1b).

As was mentioned above, not all the waves may propagate in the periodic medium due to the presence of forbidden bands (see figure 4). Particularly, the part of TR emitted in the forward direction by the particle flying into the laminated medium (main part of TR), that fits the range of forbidden bands of laminated medium ($B_1, B_2$ in figure 4), also cannot propagate. This part of TR shall be reflected by the laminated medium to vacuum. Such a throwing over of particle TR shall be followed by the appearance of appropriate maxima in the spectral-angular distribution $I_{2,v \to -v}(\omega, \theta_2)$ of radiation energy in backward direction (to vacuum).

Here (in figure 5), the spectral-angular distribution $I_{2,v \to -v}(\omega, \theta_2)$ of the energy of backward radiation from $30 MeV$ electron in flight from vacuum to laminated medium is for the same values of parameters as those in figure 3 (TR range).

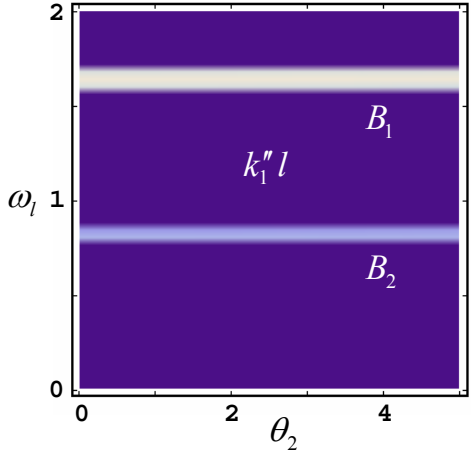 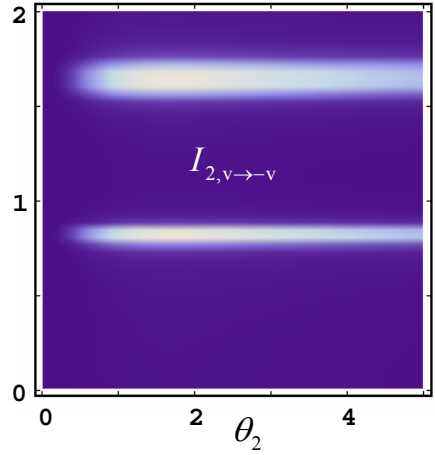

**Figure 4.** The imaginary part of quasi-wave number $k_1(\omega, u)$ (see (11)) versus $\omega$ и $\theta_2$ (see (9)) for a laminated medium with non-harmonic variations of $\varepsilon_1 = \varepsilon_1'$ (refer to figure 2, $\mu = 1$). The light-colored areas of the figure correspond to forbidden bands $B_1, B_2$ with $k_1'' > 0$ (the lighter the area, the higher is the value of $k_1''$).

**Figure 5.** Spectral-angular distribution of the energy of backward radiation from electron in flight from vacuum to the laminated medium. Calculations are based on the model of non-harmonic variations of $\varepsilon_1$. The values of system parameters are the same as those in figure 3. The value of $I_{2,v\to-v}$ is the higher, the brighter is the point representing this value.

As was to be expected, to two forbidden bands $B_1, B_2$ in figure 4 there almost exactly correspond two ranges of maxima of $I_{2,v\to-v}(\omega,\theta_2)$ in figure 5. For both the maxima

$$I_{2,v\to-v}(\omega,\theta_2) \cong I_{2,v\to-v}^{(i)}(\theta_2), \quad \text{when} \quad (\omega,\theta_2) \in B_i \quad \text{and} \quad i = 1,2. \quad (20)$$

The plots of functions $I_{2,v\to-v}^{(i)}(\theta_2)$ are shown in figure 6. It is noteworthy that as the absorption in the laminated medium decreases

$$I_{2,v\to-v}^{(i)}(\theta_2) \to I_2^{(0)}(\theta_2) \quad \text{at} \quad \varepsilon_1''/\varepsilon_1' \to 0 \quad \text{and} \quad i = 1,2. \quad (21)$$

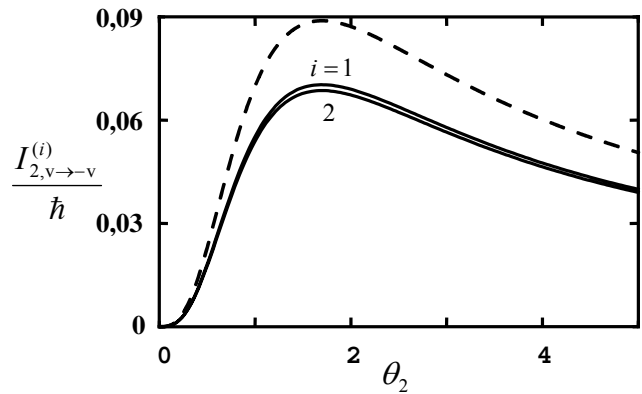

**Figure 6.** Dependence of $I_{2,v\to-v}^{(i)}(\theta_2)$ functions (see (20)) on angle $\theta_2$ for an electron in flight from vacuum to a periodical non-dispersive medium. The parameters of system are the same, as those in figure 3. The dashed curve is the limit of $I_{2,v\to-v}^{(i)}(\theta_2)$ in the absence of absorption.

Here $I_2^{(0)}(\theta_2)$ (dashed curve in figure 6) is the well known [5-7] angular distribution of TR in the forward direction for a particle flying to vacuum out of non-absorbing and non-dispersive medium with $\varepsilon' = 1.5$, $\varepsilon'' = 0$ and $\mu = 1$ (confer with (16)). Expressions (20), (21) are confirmed by our numerical calculations. In reality $I_{2,v\to-v}^{(i)} \cong I_2^{(0)}$ already for $\varepsilon_1''/\varepsilon_1' = 0.001$. So, the inequality $I_{2,v\to-v}^{(i)} < I_2^{(0)}$ in figure 6 is due to the absorption of radiation by the laminated medium on the one side, and the presence of formation zone for the radiation reflected by the periodical structure on the other side.

## 5. Conclusions

In the present work we have investigated the radiation from a charged particle flying out of a semi-infinite laminated medium to vacuum and back, - from vacuum to the laminated medium. In both the cases expressions were obtained for the spectral-angular distribution of radiation energy emitted to vacuum. In these expressions there were no limitations on the amplitude and variation profile of the dielectric permittivity $\varepsilon_1(z)$ of the laminated medium.

The results of appropriate numerical calculations indicate that:
- The spectral-angular distribution of the energy of parametric radiation from a relativistic particle along any separate branch of distribution is determined by the amplitude and profile of the variations of $\varepsilon_1(z)$.
- The spectral distribution of the energy of transition radiation of particle may be superimposed by the peaks from the diffracted beams of parametric radiation.
- Some part of transition radiation that is emitted in the forward direction at the flight of particle from vacuum to the laminated medium and which fall within forbidden bands of the periodical structure (except for the absorbed part of radiation) is completely thrown over by this structure back to vacuum.

The obtained results may be of use for controlling the parameters of charged particle radiation in laminated media, as well as for electron probing of periodical structures, for instance, of photonic crystals.

**Acknowledgments**

L Sh Grigoryan is thankful to X Artru, K A Ispirian, A P Potylitsyn, N F Shulga and A A Tishchenko for valuable comments. This work was supported in part by the Grant No.077 from Ministry of Education and Science of RA.